\documentstyle[preprint,pra,aps]{revtex}
\tightenlines
\begin{document}
\draft
\title{Calculations of the Relativistic Effects in Many-Electron Atoms
and Space-Time Variation of Fundamental Constants}

\author{ V. A. Dzuba$^*$, V. V. Flambaum and J. K. Webb}
\address{School of Physics, University of New South Wales,
	Sydney, 2052, Australia}
\date{\today}
\maketitle
\begin{abstract}
Theories unifying gravity and other interactions suggest the
possibility of spatial and temporal variation of physical
``constants'' in the Universe.  Detection of high-redshift absorption
systems intersecting the sight lines towards distant quasars provide a
powerful tool for measuring these variations.  We have previously
demonstrated that high sensitivity to the variation of the fine
structure constant $\alpha$ can be obtained by comparing spectra of
heavy and light atoms (or molecules).  Here we describe new
calculations for a range of atoms and ions, most of which are commonly
detected in quasar spectra: Fe~II, Mg~II, Mg~I, C~II, C~IV, N~V, O~I,
Al~III, Si~II, Si~IV, Ca~I, Ca~II, Cr~II, Mn~II, Zn~II, Ge~II 
(see the results in Table 3). 
The combination of Fe~II and Mg~II, for which accurate laboratory
frequencies exist, have already been used to constrain $\alpha$
variations.  To use other atoms and ions, accurate laboratory values
of frequencies of the strong E1-transitions from the ground states are
required. We wish to draw the attention of atomic experimentalists to
this important problem.

We also discuss a mechanism which can lead to a greatly enhanced
sensitivity for placing constraints on variation on fundamental
constants.  Calculations have been performed for Hg~II, Yb~II, Ca~I
and Sr~II where there are optical transitions with the very small
natural widths, and for hyperfine transition in Cs~I and Hg~II.
\end{abstract}


\section{Introduction}

Possible variations of the fundamental physical constants in the
expanding Universe are currently of particular interest because of the
implications from unified theories, such as string theory and
M-theory, that additional compact dimensions of space may exist.  The
``constants'' seen in our three-dimensional subspace of the theory
will vary at the same rate as any change occurring in the scale
lengths of the extra compact dimensions (see,
e.g. \cite{Marciano84,Barrow87,Damour94}).  Gas clouds which intersect
the sightlines towards distant quasars produce absorption lines.
These absorption systems present ideal laboratories in which to search
for any temporal or spatial variation of fundamental constants by
comparing the observed atomic spectra from the distant objects with
laboratory spectra (see, e.g. \cite{Potechin95} and references
therein).

The energy scale of atomic spectra is given by the atomic unit
$\frac{me^4}{\hbar^2}$. In the non-relativistic limit, all atomic
spectra are proportional to this constant and analyses of quasar
spectra cannot detect any change of the fundamental constants.
Indeed, any change in the atomic unit will be absorbed in the
determination of the red shift parameter $z$ ($1 + z =
\frac{\omega}{\omega '}$, $\omega '$ is the red-shifted frequency of
the atomic transition and $\omega$ is the laboratory value).  However,
any change of the fundamental constants can be found by measuring the
relative size of relativistic corrections, which are proportional to
$\alpha^2$, where $\alpha = e^2/\hbar c$ is the fine structure
constant \cite{other}.

It is natural to search for any changes in $\alpha$ using measurements
of the spin-orbit splitting within a specific fine-structure
multiplet, and indeed this method has been applied to quasar spectra
by several groups. However, whilst this method is appealing through
its simplicity, it is possible to improve on its efficiency.  Also, it
may even give incorrect results, since 
corrections of higher order than $\alpha^2$ are ignored.
An order of magnitude sensitivity gain can be achieved by
comparing transition frequencies of heavy and light atoms (or
molecules) \cite{Dzuba98,Webb}.  In this paper we extend that previous
work, presenting the results of calculations of the dependence of the
transition frequencies on $\alpha$ for many atoms and ions where data
exist for quasar spectra (see Table \ref{main}).

Other possibilities for measuring changes in $\alpha$ include
comparisons of different optical transitions, such as $s-p$ and $p-d$,
in the same atom or molecule, or comparisons of microwave transitions
in molecules which contain rotational and hyperfine intervals.  We
have also calculated the dependence on $\alpha$ of some atomic
microwave and optical frequency standards which could be used for
laboratory searches for $\alpha$ variations.  For example, one can
compare the Hg~II E2 transition $\omega=$ 35514 cm$^{-1}$ with any
narrow line of another atomic or molecular transition having
approximately the same frequency.  Small frequency differences (which
do not require very precise {\it absolute} calibration) can be
measured very accurately.  The mercury frequency has a very large
relativistic shift which also has a negative sign (usually the shift
is positive).  Therefore, this frequency difference has a very strong
$\alpha$-dependence.

Finally, there is an interesting possibility for studying transitions
between ``accidentally'' degenerate levels in the same atom or molecule. 
There are several practically degenerate levels of different electron 
configurations in rare-earth and actinide atoms. Of course, there are many
more possibilities in molecules where there are vibrational and
rotational structures.  The relativistic corrections to the different
energy levels are different and can exceed the very small frequency
corresponding to the transition between ``degenerate'' states by many
orders of magnitude, i.e. a tiny variation of $\alpha$ can change the
frequency significantly.  Also, there is an interesting dependence on
the nucleon mass if the rotational, vibrational and hyperfine
structures are involved. In this case one can measure time dependence
of the nucleon mass which is a function of the strong interaction
constant.  The main problem here is to find a narrow-width transition.

Note that we present all results in this paper assuming that the atomic
unit of energy $\frac{me^4}{\hbar^2}$ is constant.

\section{Theory}

\subsection{Semi-empirical estimations}

Let us start our calculations using simple analytical estimates of the
relativistic effects in transition frequencies.  First consider the
relativistic corrections to the frequency of an atomic transition in a
hydrogen-like atom. The relativistic correction to the energy level is
given by (see, e.g. \cite{Drell})
\begin{eqnarray}
	\Delta_n = - \frac{me^4Z^2}{2\hbar^2}\frac{(Z\alpha)^2}{n^3}
	(\frac{1}{j + 1/2} - \frac{3}{4n}),
\label{rel}
\end{eqnarray}
where $Z$ is the nuclear charge, $n$ is the principal quantum number
and $j$ is the total electron angular momentum. This value of the
relativistic correction can be obtained as an expectation value
$\langle V \rangle$ of the relativistic perturbation $V$, which is
only large in the vicinity of the nucleus. Therefore, the relativistic
correction $\Delta$ is proportional to the electron density near the
nucleus $|\Psi(r<\frac{a}{Z})|^2 \propto \frac{Z^3}{n^3 a^3}$ ($a$ is
the Bohr radius, $\frac{a}{Z}$ is the size of the hydrogen-like ion).
For an external electron in a many-electron atom or ion the electron
density near the nucleus is given by the formula (see,
e.g. \cite{Sobelman}) obtained in the semi-classical approximation ($n
\gg 1$)
\begin{eqnarray}
	|\Psi(r < \frac{a}{Z})|^2 \propto \frac{Z^2_a Z}{\nu^3 a^3},
\label{dencity}
\end{eqnarray}
where $Z_a$ is the charge ``seen'' by the external electron outside
the atom, i.e. $Z_a$ = 1 for neutral atoms, $Z_a$ = 2 for singly
charged ions, etc.; $\nu$ is the effective principal quantum number,
defined by $E_n = - \frac{me^4}{2\hbar^2}\frac{Z_a^2}{\nu^2}$, where
$E_n$ is the energy of the electron. For hydrogen-like ions $\nu = n,
Z_a = Z$.  Thus, to find the single-particle relativistic correction,
we should multiply $\Delta$ in Eq. (\ref{rel}) by the ratio of
$|\Psi(r<\frac{a}{Z})|^2$ in the multi-electron ion and hydrogen-like
ion. The result is
\begin{eqnarray}
	\Delta_n = - \frac{me^4 Z_a^2}{2\hbar^2}\frac{(Z\alpha)^2}
	{\nu^3} \left[\frac{1}{j+1/2}-\frac{Z_a}{Z\nu}(1-\frac{Z_a}{4Z})\right]
 \simeq E_n \frac{(Z\alpha)^2}{\nu (j+1/2)}.
\label{rel1}
\end{eqnarray}
The second term in the square brackets is presented to provide a
continuous transition from the hydrogen-like ion  Eq. (\ref{rel}) to
the multi-electron ion  Eq. (\ref{rel1}). In multi-electron ions  ($ Z \gg 
Z_a$) this term is, in fact, a rough  estimate based on the direct calculation 
of $\langle V \rangle$.  We should neglect this small term  since there
are more important many-body corrections.

We see that the relativistic correction is largest for the $s_{1/2}$ and 
$p_{1/2}$ states, where $j = 1/2$. The fine structure splitting is given by
\begin{eqnarray}
\Delta_{ls} = E(p_{3/2}) - E(p_{1/2}) \simeq - \Delta(p_{1/2})/2 \simeq 
 - \Delta(p_{3/2}).
\label{rel2}
\end{eqnarray}

In quasar absorption spectra, transitions from the ground state have
been observed. Therefore, it is important to understand how the
frequencies of these transitions are affected by relativistic effects.
The fine splitting in excited states is smaller then the relativistic
correction in the ground state, since the density of the excited
electron near the nucleus is smaller. As a result, the fine splitting
of the $E1$-transition from the ground state (e.g., $s -p$) is
substantially smaller than the absolute shift of the frequency of the
$s-p$ transition. The mean energy of the $p$-electron is defined as
\begin{eqnarray}
	E(p) = \frac{2}{3}E(p_{3/2}) + \frac{1}{3}E(p_{1/2}) \simeq
	E_n(p) - \frac{4}{3}\Delta_{ls},
\label{rel3}
\end{eqnarray}
where $E_n$ is the non-relativistic energy. Therefore, the relativistic 
shift of the mean $s-p$ transition frequency is given by
\begin{eqnarray}
	\Delta(p-s) \simeq - \frac{4}{3}\Delta_{ls} - \Delta(s_{1/2}).
\label{rel4}
\end{eqnarray}
The formulae (\ref{rel1}) - (\ref{rel4}) do not take into account
many-body effects. For example, relativistic corrections change the
self-consistent atomic potential.  The many-body calculations
discussed below show that the relativistic energy shift in atoms with
one external electron can be approximately described by the following
equation
\begin{eqnarray}
	\Delta_n = \frac{E_n}{\nu}(Z\alpha)^2 \left[\frac{1}{j+1/2} -
	C(Z,j,l)\right],
\label{rel5}
\end{eqnarray}
where $C(Z,j,l)$ is different for different atoms and partial waves
but does not depend on the principal quantum number. In many cases
$C(Z,j,l) \simeq 0.6$, although noticeable deviations from this
value are possible. 

It is easy to explain the sign of the many-body effect.  The
relativistic single-particle correction increases the attraction of an
electron to the nucleus and makes the radius of the electron cloud
smaller. As a result, the direct Hartree-Fock potential, which is the
nuclear potential screened by core electrons, becomes smaller at short
distances. This decreases the binding energy of a valence electron.
Therefore, many-body effects have an opposite sign to the direct
effect.

It is also easy to see why $C(Z,j,l)$ does not depend on the energy of
the valence electron.  The effect of a potential change on short
distances on the binding energy of an external electron is
proportional to the density of this electron in the vicinity of the
nucleus which in turn is proportional to $1/\nu^3$ (see
eq. (\ref{dencity})).  On the other hand, the direct relativistic
effect is also proportional to $1/\nu^3$ for the same reason (see
eqs. (\ref{dencity},\ref{rel1})).  Therefore the ratio of two effects
which is $C(Z,j,l)/(2j+1)$ is practically independent of the energy of
the external electron.

The accurate value of the relativistic shift can only be obtained from
many-body calculations. However, one can assume that $C(Z,j,l) = 0.6$
and use (\ref{rel5}) for rough estimates of relativistic
corrections. This usually gives better results than a single-particle
estimate in eq. (\ref{rel1}). For example, many-body calculations show
that, as a rule, the relativistic correction to the energy is negative
for $s_{1/2}$ and $p_{1/2}$ states and positive for other states. This
behavior is reproduced by (\ref{rel5}) but not by (\ref{rel1}).  Apart
from that, formula (\ref{rel1}) suggests that the correction is
largest for $s_{1/2}$ and $p_{1/2}$ states and rapidly decreases with
j while many-body calculations show that the correction for $d$-states
is sometimes bigger than for $p$-states. This is again reproduced by
(\ref{rel5}) where there is a strong cancelation between two terms in
the case of $p$-states.  Note that this complex behavior of the
relativistic effects cannot be explained in terms of single-electron
density at the origin and should be attributed to many-body effects.

Let us see now how eqs. (\ref{rel2})-(\ref{rel4}) will be modified if
the many-body correction $C(Z,j,l)$ is included.  Assuming the same
value of $C(Z,j,l)$ for $p_{1/2}$ and $p_{3/2}$ states we have for the
fine splitting
\begin{eqnarray}
	\Delta_{ls} = E(p_{3/2}) - E(p_{1/2}) \simeq 
	- \frac{1}{2} E_n \frac{(Z\alpha)^2}{\nu (j+1/2)}.
\label{rel6}
\end{eqnarray}
For the mean energy of the $p$-electron instead of (\ref{rel3}) we have
\begin{eqnarray}
	E(p) = 	E_n(p) + (2C - \frac{4}{3})\Delta_{ls}.
\label{rel7}
\end{eqnarray}
With $C = 0.6$ the second term in (\ref{rel7}) is numerically small.
Hence, relativistic corrections move the $p_{1/2}$ and $p_{3/2}$
states in opposite directions, leaving the mean energy almost
unchanged.  Therefore, the relativistic shift of the mean $s - p$
transition is mostly given by the energy shift of the $s$-state
\begin{eqnarray}
	\Delta(p-s) \simeq - \Delta(s_{1/2}).
\label{rel8}
\end{eqnarray}
The relative sizes of relativistic corrections are proportional to
$Z^2$, so they are small in light atoms. Therefore, we can constrain
changes in $\alpha$ by comparing transition frequencies in heavy and
light atoms. We stress that the most accurate and effective procedure
includes all relativistic corrections and the analysis of all
available lines (rather then the fine splitting within one multiplet
only).  We have not discussed here contribution of the Breit
relativistic correction to the electron-electron interaction. It is
not enhanced by the factor $Z^2$ and is much smaller then the
contribution from spin-orbit interaction. Our numerical calculations
have demonstrated that the contribution of the Breit interaction to
the frequency of s-p transition $\Delta(p-s)$ is indeed negligible.

\subsection{Relativistic many-body calculations}

Accurate calculations of relativistic effects in atoms have been done
using many-body theory which includes electron-electron correlations.
We used a correlation-potential method \cite{Dzuba87} for atoms with
one external electron above closed shells and a combined configuration
interaction and many-body perturbation theory method \cite{Kozlov} for
atoms with several valence electrons. A relativistic Hartree-Fock
(RHF) Hamiltonian was used in both cases to generate a complete set of
single-electron orbitals.  The value of the relativistic corrections
were obtained by repeating the calculations for different values of
$\alpha$ in the RHF Hamiltonian.

The form of the the single-electron wave function we use explicitly
includes a dependence on the fine structure constant $\alpha$

\begin{eqnarray}
\psi({\bf r})_{njlm} = \frac{1}{r}
 \left( \begin{array}{c}
f(r)_n \Omega({\bf r}/r)_{jlm} \\ i 
\alpha g(r)_n \tilde{\Omega}({\bf r}/r)_{jlm}
\end{array}
\right).
\label{psi}
\end{eqnarray}
This leads to the following form of the RHF equations
\begin{eqnarray}\label{RHF}
f'_n(r)+\frac{\kappa_n}{r}f_n(r)-[2+\alpha^2(\epsilon_n - \hat V)]g_n(r) = 0\\
g'_n(r)-\frac{\kappa_n}{r}g_n(r)+(\epsilon_n - \hat V)f_n(r) = 0 \nonumber ,
\end{eqnarray}
where $\kappa = (-1)^{l+j+1/2}(j+1/2)$ and $V$ is Hartree-Fock potential:
\begin{eqnarray}
\hat V f = V_d(r)f(r) - \int V_{exch}(r,r')f(r')dr' .
\label{hfpot}
\end{eqnarray}
The non-relativistic limit can be achieved by reducing the value of
$\alpha$ to $\alpha = 0$.

For atoms or ions with one external electron above closed shells, the
calculations begin with the $V^{N-1}$ approximation.  A Hartree-Fock
procedure is carried out initially for a closed shell ion with the
external electron removed. When convergence for the core is achieved,
the states of the external electron are calculated in the field of the
frozen core. The main reason for this approach is the simplicity of
the perturbation theory for calculation of the correlation
corrections. It is well known that correlations are very important in
many-electron atoms and should be included into the calculations to
obtain accurate results. We do this by means of many-body perturbation
theory (MBPT) and a correlation potential method \cite{Dzuba87}.

The correlation potential $\hat \Sigma$ is defined as an operator
which gives a correlation correction to the energy (ionization
potential) of the valence electron
\begin{equation}
	\delta \epsilon_a = \langle a | \hat \Sigma | a \rangle ,
\label{Sigma}
\end{equation}
The expectation value here is taken over the single-particle wave
function of the external (valence) electron. Thus, $\hat \Sigma$ is
another non-local operator which can be included in (\ref{hfpot}) by
redefining the non-local potential
\begin{equation}
V_{exch}(r,r') \rightarrow V_{exch}(r,r') + \Sigma(r,r').
\label{non-local}
\end{equation}
Single-electron states of a valence electron calculated in a non-local
potential (\ref{non-local}) are often called Brueckner orbitals.

MBPT is used to calculate $\Sigma$. Perturbation expansion starts from
the second order in residual Coulomb interaction.  Dominating
higher-order correlations can also be included by using the technique
developed in Refs. \cite{Dzuba89}. However we found that for the few
electron ions such as C~IV and Si~IV, second-order results are already
of very high accuracy (see Table \ref{CSi}). Therefore, we decided
that the accurate calculations of higher order corrections are not
needed.
 
For some other atoms we introduce fitting parameters $f_v$ into the
expression for the non-local potential in (\ref{hfpot}) to simulate
the effect of higher-order correlations: $V_{non-local} = V_{exch} +
f_v \Sigma$, where $v = s,p$ or $d$.  The values for $f_s, f_p$ and
$f_d$ are chosen to fit the experimental data for laboratory value
$\alpha = 1/137.036$. In all cases the values of $f_v$ are close to
unity.  The same $f_v$ have been used for calculations with varying
$\alpha$.  This procedure works well because the accuracy of the
results in the second order is already good and only a small
correction is introduced by the fitting parameters.

For atoms with more then one external electron we use the combination
of the configuration interaction method with many-body perturbation
theory \cite{Kozlov}.
\begin{enumerate}
\item As for single-electron-above-closed-shells atoms, we start
calculations from the RHF method in $V^{N-1}$ approximation.  However,
this starting approximation does not correspond to a closed-shell
system and needs to be further specified.  We do this in a very simple
way. The contribution of an open shell to the Hartree-Fock potential
is calculated using the complete shell potential multiplied by the
``occupation'' factor $n/(2j + 1)$, where $n$ is the actual number of
electrons on that shell and $j$ is the total single-electron momentum.
Single-electron basis states are calculated in this Hartree-Fock
potential with one external electron removed ($V^{N-1}$ potential).
\item All basis states are divided into core states and valence
states. Core states are frozen and included into calculations only via
the effective potential of the core. Valence states are used as a
basis for the configuration interaction method. Note that the
definition of the core at this stage does not necessarily coincide
with the core in RHF calculations. For example, for Fe~II $3d_{3/2}$
and $3d_{5/2}$ states are core states in the RHF procedure. But these
states are also included into the configuration interaction, so they
are valence states in the CI calculations.
\item The effective Hamiltonian of the CI method is constructed.  To
include correlations between core and valence electrons we modify the
effective Hamiltonian of the standard CI method by adding an extra
operator $\Sigma$
\[
	\hat H^{CI}_{eff} \rightarrow H^{CI}_{eff} + \hat \Sigma.
\]
The operator $\Sigma$ consists of two parts. $\Sigma_1$ is the
one-electron operator which describes the correlation interaction
between a valence electron and the core. $\Sigma_1$ is very similar to
the correlation potential $\Sigma$ in (\ref{non-local}) which we used
for atoms with one external electron above closed shells.  $\Sigma_2$
is a two-particle operator which describes the effects of screening of
the Coulomb interaction between valence electrons by the core
electrons.

It has been demonstrated in Refs. \cite{Kozlov} that the core-valence
correlations are very important and usually dominate over correlations
between valence electrons. Thus, it is more important to include
$\Sigma$ than to achieve the completeness of the basis in the CI
calculations.

\item The standard CI technique is used to diagonalize the matrix of
the effective CI Hamiltonian and obtain many-electron energies and
wave functions.
\end{enumerate}
We need to apply this method to the atoms with many electrons in open
shell. This can make configuration space and computation time very
large.  However, we do not need very high accuracy in the present
calculations.  Therefore, we have made a few simplifications to the
method compared to Ref. \cite{Kozlov}.  The effect of screening
($\Sigma_2$) is usually much smaller than one-electron correlations
with the core ($\Sigma_1$) and we neglect it.  We also neglect
subtracting diagrams in $\Sigma_1$ (see Ref. \cite{Kozlov}) because
there must be strong cancelations between screening and subtracting
diagrams.  Finally, we use a relatively small basis set with 3 to 4
single-electron basis states of each symmetry. This usually leads to a
few hundred configurations. This small basis is not complete to high
precision. However the results are reasonably good because
correlations with the core are included (see Table \ref{CrII}).  To
simulate the effect of incompleteness of the basis and of the omitted
diagrams we introduce fitting parameters for $\Sigma$, similar to what
we did for atoms with one external electron.  The results for Cr~II
presented in Table \ref{CrII} illustrate the effects of core-valence
correlations, correlations between valence electrons and the effect of
fitting. Note that only one fitting parameter was used to fit all
energy levels.

There are two contributions to the relativistic energy shift.  The
first is the direct relativistic correction to the energy of a valence
electron in the Hartree-Fock-Dirac equations (\ref{RHF}). This
correction can be found be varying $\alpha$ in eq. (\ref{RHF}) with
fixed potential. There is also an indirect relativistic correction
which appears because of the change of the core potential (including
correlation potential) due to the relativistic effects in the core.
Neither the correlation potential $\Sigma$, nor the standard
Hartree-Fock potential depend on $\alpha$ explicitly. This dependence
appears via the basis set of single-electron wave functions used to
calculate both potentials since these wave functions had been obtained
by solving Dirac-like equations (\ref{RHF}).  Full-scale calculations,
repeated for different values of $\alpha$, were necessary to reveal
this implicit relativistic behavior. The value of the indirect effect
is not small and in some cases exceeds several times the value of the
direct one.  This indirect relativistic effect is essentially a
many-body effect. It is responsible for the failure of the
single-particle formula (\ref{rel1}) which is unable to reproduce the
value of the relativistic correction accurately.

\section{Results and discussion}

To find the dependence of frequencies on $\alpha$ we use the following 
formula for the energy levels within one fine-structure multiplet:
\begin{eqnarray}
	E = E_0 + Q_1 ((\frac{\alpha}{\alpha_l})^2-1) +
	         Q_2 ((\frac{\alpha}{\alpha_l})^4-1) +
	K_1 ({\bf LS})(\frac{\alpha}{\alpha_l})^2 +
	K_2 ({\bf LS})^2 (\frac{\alpha}{\alpha_l})^4.
\label{ls}
\end{eqnarray}
Here $E_0, Q1$ and $Q_2$ describe the position of the configuration
center, $K_1$ and $K_2$ describe the level splitting within one
configuration, $L$ is the total orbital angular momentum, $S$ is the
total electron spin and $\alpha_l$ is the laboratory value of
$\alpha$.  We introduce an $({\bf LS})^2$ term to describe deviations
from the Lande interval rule. There are two sources of the $({\bf
LS})^2$ term: the second order in the spin-orbit interaction ($ \sim
(Z \alpha)^4$) and the first order in the Breit interaction ($ \sim
\alpha^2 = 5.3 \times 10^{-5}$).

The second-order spin-orbit interaction is larger for heavy atoms
where we actually need to introduce the $({\bf LS})^2$ term.
Therefore, we first fitted the experimental fine structure intervals
to find $K_1$ and $K_2$ (numerical calculations give close values of
$K_1$ and $K_2$). Then we used numerical calculations for $\alpha =
\sqrt{7/8}\alpha_l$ and $\alpha = \sqrt{3/4}\alpha_l$ to find the
dependence of the configuration center on $\alpha$ (coefficients $Q_1$
and $Q_2$).  It is convenient to represent the final results in the
form
\begin{eqnarray}
	\omega = \omega_0 + q_1 x + q_2 y,
\label{omegaxy}
\end{eqnarray}
where $q_1 = Q_1 + K_1 ({\bf LS}), q_2 = Q_2 + K_2 ({\bf LS})^2$,
$x = (\frac{\alpha}{\alpha_l})^2 - 1, y = (\frac{\alpha}{\alpha_l})^4 - 1$
and $\omega_0 = E_0 + K_1 ({\bf LS}) + K_2 ({\bf LS})^2 $ is an
experimental energy of a particular state of the fine structure
multiplet.  The parameters $\omega_0, q_1$ and $q_2$ for E1
transitions for many atoms and ions of astrophysical interest are
presented in Table \ref{main}.

One can use these data to fit absorption systems in quasar spectra in
order to measure or place upper limits on any variation of $\alpha$.
The maximum theoretical sensitivity comes from comparing the spectra
of Fe~II and Cr~II since relativistic effects in both ions are large
and have opposite sign. The effect here is about 20 times larger than
the fine splitting for each of the ions.

An analysis of the theoretical data reveals some interesting tendencies
in the behavior of the relativistic corrections apart from being
proportional to $Z^2$:
\begin{itemize}
\item Within a series of one-electron orbitals of a given symmetry,
the relativistic energy shift is largest for the lowest orbital and
decreases for orbitals of higher energies. This trend is supported by
the semi-empirical consideration presented in section 1.  The higher
electron density in the vicinity of the nucleus gives larger
relativistic effects. This also explains why the relativistic shift of
the ground state energy is bigger when ionization potential is larger.
\item When $\alpha$ is changing towards its non-relativistic limit
$\alpha = 0$, one-electron energies of $s$ and $p_{1/2}$ states move up
while energies of $p_{3/2}$ and $d$ states move down. The relativistic
shifts of $s$ and $d$ states tend to be large while the energy shifts of
$p$ states are relatively small. Note that the single-particle consideration
suggests that the relativistic shifts are large for $s$ and $p_{1/2}$
states and all energies move up when $\alpha$ is decreasing (see
formulae (\ref{rel1} - \ref{rel3}).
Accurate calculations give different behaviors due to indirect relativistic 
effects: relativistic corrections to the core orbitals change the electronic
potential which in turn shifts the energy of the external electron.
This effect is neglected in a naive single-particle formulae. 
The direct effect dominates over the core change effect for
$s$-states.  For $p$-states these two effects are close in magnitude
but opposite in sign.  The core change effect dominates for
$d$-states. Thus, introduction of $C(Z,j,l)$ in eq. (\ref{rel6})
effectively takes into account the core change effect.
\end{itemize}
These tendencies are illustrated by the results presented in Table
\ref{main}.  For example, the relativistic shifts of frequencies of
$E1$ transitions are negative in Cr~II and Ge~II and positive for
other atoms. This is because these transitions correspond to a $d - p$
one-electron transition in the case of Cr~II and to a $p - s$
transition for Ge~II while for the other atoms the transitions are of
the $s - p$ type.  The relativistic correction for Ge~II is relatively
small because the corresponding transition may be described as the
transition from the ground $p$ orbital to excited $s$ orbital and
because the relativistic corrections are small for both $p$ states and
excited states.  Another example is the sharp increase of the
relativistic effect from Ca~II to Cr~II. This is due to the coherent
effect of two factors: bigger Z and bigger ionization potential for Cr~II 
as compared to Ca~II.

As is apparent from the analysis above one should expect the biggest
relativistic shift for the $s - d$ (or $d - s$) transitions in heavy
atoms.  These transitions are not observed in quasar absorption
systems but may be suitable for laboratory experiments. The natural
line width for these transitions is very small. Futhermore, very
precise measurements of the frequencies of many such transitions
already exist since they are used as atomic optical frequency
standards. We present in Table \ref{tabFS} relativistic shifts of
frequencies of some atomic transitions which are used or proposed as
optical frequency standards. These include the strongly forbidden $E1$
transition in Ca~I \cite{Ca} and $E2$ transitions in Sr~II \cite{Sr},
Yb~II \cite{Yb} and Hg~II \cite{Bergquist}.  There are many other
atoms which are being studied as possible frequency standards but
which are not included in the table. These include Mg~I\cite{Beveriny},
In~II \cite{Peik}, Xe~I \cite{Walhout}, Ag~I \cite{Larkins}, etc.
Note that the biggest relativistic effect is in Hg~II.  This is
because of the $d - s$ transition and a high value of $Z$.  This makes
Hg~II the most interesting candidate for a laboratory search for
$\alpha$ variations.

The following limit of $\alpha$ variation was found in Ref. \cite{Webb}:
\begin{equation}
\begin{array}{ll}
\frac{\left\langle \dot{\alpha}\right\rangle}{\alpha} = +2.6 \pm 5.2
\times 10^{-16}
\end{array}
\label{timederiv}
\end{equation}
Assuming $\dot\alpha/\alpha = 10^{-15}$ one can get for the 
$5d^{10}6s \ ^2S_{1/2} - 5d^96s^{2 \ 2}D_{5/2}$ transition in Hg~II
\[
\dot\omega = 3 \mbox{Hz yr}^{-1},
\]
which should be compared with the natural linewidth limit of 1.8 Hz
\cite{Bergquist}.

There are also ongoing laboratory searches for variations of $\alpha$
using microwave atomic frequency standards (atomic clocks) (see,
e.g. \cite{Prestage}). There are a number of microwave frequency
standards which use ground state hyperfine structure (hfs) intervals
of atoms or ions (see, e.g. review \cite{Fisk}).  These include hfs of
Rb, Cd$^+$, Cs,
Ba$^+$, Yb$^+$, Hg$^+$ \cite{Fisk}.
Here again the biggest relativistic effects are in the Hg$^+$ ion.
It is convenient to present the $\alpha$-dependence of hfs constants
in a form similar to (\ref{omegaxy})
\begin{eqnarray}
	A_{hfs} = (\frac{\alpha}{\alpha_l})^2 (A_0 + q x),
\label{hfs}
\end{eqnarray}
where $A_0$ is the hfs constant for $\alpha = \alpha_l$. 
Its value was recently measured to very high precision:
\[
	A_0 (\mbox{Hg}^+) = 40507347996.84159(14)(41) \mbox{Hz}
	\cite{Berkeland}.
\]
Many-body calculations similar to the calculations of energies 
described above show that $q = 40500$ MHz for Hg~II and $q = 956$ MHz
for Cs~I. Note that $A_0$($^{133}$Cs) = 2298157943 Hz. This is the exact
value because the frequency 9192631770 Hz of the $6s~F=3 - 6s~F=4$ hfs 
transition in $^{133}$Cs, which is equal to $4A_0(6s)$, is used 
as a definition of the metric second.

Relative drifts in rates of atomic clocks based on Hg~II and Cs will
be
\[
	\frac{d}{dt}\ln \frac{A(\mbox{Hg})}{A(\mbox{Cs})} = 
	( \frac{2q}{A_0}(\mbox{Hg}^+) -
	\frac{2q}{A_0}(\mbox{Cs})) \frac{\dot \alpha}{\alpha} =
	(2.30 - 0.83)  \frac{\dot \alpha}{\alpha} =
	1.47 \frac{\dot \alpha}{\alpha}.
\]
This is in good agreement with estimates based on the Fermi-Segre
formula \cite{Prestage}.  The best sensitivity of various clock rate
comparisons can be achieved when Hg$^+$ clocks are compared with an H
maser \cite{Prestage} (parameter q = 0 for H). Assuming again
$\dot\alpha/\alpha = 10^{-15}$ yr$^{-1}$ we will get $2.3 \times
10^{-15}$yr$^{-1}$ for the frequency rate shift between H maser and
Hg$^+$ clocks. Note that the ratio of hyperfine structure constants is
also sensitive to the variation of nuclear magnetic $g$-factors which
may appear due to variation of strong interaction.

One more interesting possibility is to use transitions between
``accidentally'' degenerate levels in the same atom. Such meta-stable
levels exist, for example, in the Dy atom: two $J=10$ opposite parity
levels $4f^{10}5d6s$ and $4f^95d^26s$ lying 19797.96 cm$^{-1}$ above
ground state. (This pair of levels was used to study parity
non-conservation in Refs. \cite{DzubaDy,Budker}). There are other
examples of ``accidentally'' degenerate levels in the rare-earth and
actinide atoms and many close levels in other heavy atoms and ions (in
the absence of degeneracy one should look for $s-d$ or $s-p$
transitions where the relativistic effects are larger).  In the case
of ``accidental'' degeneracy, the contribution of the relativistic
correction to the frequency of the $E1$ transition in a heavy atom
($\sim 1000 cm^{-1}$) is compensated by the difference in the Coulomb
interaction energies of the two configurations.  However, if $\alpha$
varies with time, this compensation will eventually disappear. Thus,
we have a correction $\sim 1000$ cm$^{-1} ((\frac{\alpha}{\alpha_l})^2
-1)$ to the very small ($<$ 0.01 cm$^{-1}$) frequency of the
transition. One can measure, for example, the time dependence of the
ratio of frequencies for transitions between the hyperfine components
of these two states.  In the case of ``accidentally'' degenerate
levels belonging to different electron terms in a molecule one can
have enhanced effects of the change of both $\alpha$ and the nucleon
mass.  In the latter case the enhancement factor is the ratio of the
vibration energy to the small frequency of the transition. The problem
in these degenerate level cases is to find a transition with a small
natural width.

\section*{Acknowledgements}

We are grateful to O. Sushkov, D. Budker and J. Vigue  for useful 
discussions.
V.V.F. is grateful to Laboratoire de Physique Quantique, Universite Paul
Sabatier for hospitality.
V.A.D. is grateful to the Physics Department of the
University of Notre Dame for hospitality.


\begin{table}
\caption{Energy levels of C~IV and Si~IV with respect to the continuous
spectrum limit (cm$^{-1}$)}
\label{CSi}
\begin{tabular}{ccccc}
 ~Ion & State & RHF & Brueckner & Experiment\tablenotemark[1]  \\
\hline
 C~IV & $2s_{1/2}$ & 519253 & 520082 & 520178 \\
      & $2p_{1/2}$ & 454054 & 455508 & 455694 \\
      & $2p_{3/2}$ & 453926 & 455337 & 455587 \\
      & $3d_{3/2}$ & 195196 & 195287 & 195298 \\
      & $3d_{5/2}$ & 195187 & 195277 & 195287 \\
Si~IV & $3s_{1/2}$ & 360613 & 363840 & 364098 \\
      & $3p_{1/2}$ & 290073 & 292514 & 292808 \\
      & $3p_{3/2}$ & 289606 & 292036 & 292348 \\
      & $3d_{3/2}$ & 201807 & 203480 & 203721 \\
      & $3d_{5/2}$ & 201807 & 203472 & 203721 \\
\end{tabular}
\tablenotetext[1]{Reference \cite{Moore}; numbers are rounded to the last 
digit before the decimal point.} 
\end{table}

\begin{table}
\caption{Ionization potential and excitation energies of Cr~II (cm$^{-1}$)}
\label{CrII}
\begin{tabular}{cccccc}
 ~State & RHF\tablenotemark[1]
        & RHF + $\Sigma$\tablenotemark[2]
        & CI + $\Sigma$\tablenotemark[3]  
        & CI + $f \times \Sigma$\tablenotemark[4]  
        & Experiment\tablenotemark[5] \\
\hline
	\multicolumn{6}{c}{Ionization potential} \\
 $3d^5    ~~^6$S$_{5/2}$  & 125889 & 141067 & 137208 & 133815 & 133060 \\
	\multicolumn{6}{c}{Excitation energies} \\
 $3d^4 4p ~~^6$F$_{3/2}$  & 40230 & 52943 & 50233 & 46777 & 46905.52 \\
 $3d^4 4p ~~^6$F$_{5/2}$  & 40388 & 53125 & 50411 & 46949 & 47040.54 \\
 $3d^4 4p ~~^6$F$_{7/2}$  & 40606 & 53377 & 50655 & 47187 & 47227.50 \\
 $3d^4 4p ~~^6$P$_{3/2}$  & 42422 & 55171 & 51902 & 48466 & 48399.19 \\
 $3d^4 4p ~~^6$P$_{5/2}$  & 42562 & 55342 & 52067 & 48621 & 48491.39 \\
 $3d^4 4p ~~^6$P$_{7/2}$  & 42758 & 55581 & 52304 & 48844 & 48632.36 \\
\end{tabular}
\tablenotetext[1]{Single-configuration approximation.}
\tablenotetext[2]{As RHF but correlations with core electrons included
in the second order.}
\tablenotetext[3]{Configuration interaction for 343 relativistic 
configurations; correlations with core are also included.}
\tablenotetext[4]{Screening parameter $f_d = 0.74$ has been set for 
correlations of $d$-valence electrons with the core to fit ionization 
potential and excitation energies.}
\tablenotetext[5]{Reference\cite{Moore}.}
\end{table}

\begin{table}
\caption{Dependence on $\alpha$ of the frequencies of the $E1$ atomic
transitions of astronomic interest  (cm$^{-1}$). Here
$\omega = \omega_0 + q_1 x + q_2 y$ where 
$x = (\frac{\alpha}{\alpha_l})^2 - 1, y = (\frac{\alpha}{\alpha_l})^4 - 1$.}
\label{main}
\begin{tabular}{cllllllrr}
 ~Z & Atom/Ion & \multicolumn{2}{c}{Ground state} & 
\multicolumn{2}{c}{Upper states} & $\omega_0$ & $q_1$ & $q_2$~~ \\
\hline
 ~6 & C~II & $2s^22p$ & $ ^2P^o_{1/2}$ & $2s2p^2$ & $ ^2D_{3/2}$ &
\dec 74932.617\tablenotemark[1] & 177 & 3~~ \\
    &      &          &                & $2s2p^2$ & $ ^2S_{1/2}$ &
\dec 96493.742\tablenotemark[1] & 171 & 3~~ \\
    &      &          &                & $2s2p^2$ & $ ^2P_{1/2}$ &
\dec 110625.1\tablenotemark[2] & 173 & -3~~ \\
    &      &          &                & $2s2p^2$ & $ ^2P_{3/2}$ &
\dec 110666.3\tablenotemark[2] & 217 & 3~~ \\

 ~6 & C~IV & $1s^22s $ & $ ^2$S$_{1/2}$ & $1s^22p $ & $ ^2$P$_{1/2}$ & 
\dec 64484.094\tablenotemark[1] & 108 & 8~~ \\
    &     &        &                    & $1s^22p $ & $ ^2$P$_{3/2}$ & 
\dec 64591.348\tablenotemark[1] & 231 & -8~~ \\

 ~7 & N~V & $2s$ & $ ^2$S$_{1/2}$ & $2p$ & $ ^2$P$_{1/2}$ &
\dec 80463.211\tablenotemark[1] & 196  & -4~~  \\
    &     &      &                & $2p$ & $ ^2$P$_{3/2}$ &
\dec 80721.906\tablenotemark[1] & 488  & 2~~  \\

 ~8 & O~I & $2p^4 $    & $ ^3$P$_2$     & $2p^33s $ & $ ^3$S$_1$ &
\dec 76794.977\tablenotemark[1] & 130 & -30~~ \\
    &     &            &                & $2p^34s $ & $ ^3$S$_1$ &
\dec 96225.055\tablenotemark[1] & 140 & -20~~ \\

 ~12 & Mg~I & $3s^2$ & $ ^1$S$_0$ & $3s3p $ & $ ^1$P$_1$ & 
\dec 35051.277\tablenotemark[3] & 106 & -10~~ \\

 ~12 & Mg~II & $3s$ & $ ^2$S$_{1/2}$ & $3p $ & $ ^2$P$_{1/2}$ & 
\dec 35669.298\tablenotemark[3] & 120 & 0~~ \\

     &       &      &                & $3p $ & $ ^2$P$_{3/2}$ & 
\dec 35760.848\tablenotemark[3] & 211 & 0~~ \\

 ~13 & Al~III & $3s$ & $ ^2$S$_{1/2}$ & $3p$ & $ ^2$P$_{1/2}$ &
\dec 53682.330\tablenotemark[1]   & 216 & 0~~ \\
     &        &      &                & $3p$ & $ ^2$P$_{3/2}$ &
\dec 53916.598\tablenotemark[1]   & 464 & 0~~ \\

 ~14 & Si~II & $3s^23p$ & $ ^2P^o_{1/2}$ & $3s3p^2$ & $ ^2D_{3/2}$ &
\dec 55309.352\tablenotemark[1] & 547 & -6~~ \\

 ~14 & Si~IV & $2p^63s $ & $ ^2$S$_{1/2}$ & $2p^63p $ & $ ^2$P$_{1/2}$ & 
\dec 71287.523\tablenotemark[1] & 362 & -8~~ \\
     &      &           &                & $2p^63p $ & $ ^2$P$_{3/2}$ & 
\dec 71748.625\tablenotemark[1] & 766 & 48~~ \\

 ~20 & Ca~I & $4s^2 $ & $ ^1$S$_0$ & $4s4p $ & $ ^1$P$_1$ & 
\dec 23652.305\tablenotemark[1] & 300 & 0~~ \\
 ~20 & Ca~II & $3p^64s $ & $ ^2$S$_{1/2}$ & $3p^64p $ & $ ^2$P$_{1/2}$ & 
\dec 25191.512\tablenotemark[1] & 192 & 16~~ \\
    & & &   & $3p^64p $ & $ ^2$P$_{3/2}$ & 
\dec 25414.427\tablenotemark[1] & 420 & 16~~ \\

 ~24 & Cr~II & $3d^5$ & $ ^6$S$_{5/2}$ & $3d^44p$ & $ ^6$F$_{3/2}$ &
\dec 46905.17\tablenotemark[4] & -1624 & -25~~ \\
     &       &        &                & $3d^44p$ & $ ^6$F$_{5/2}$ &
\dec 47040.35\tablenotemark[4] & -1493 & -21~~ \\
     &       &        &                & $3d^44p$ & $ ^6$F$_{7/2}$ &
\dec 47227.24\tablenotemark[4] & -1309 & -18~~ \\
     &       &        &                & $3d^44p$ & $ ^6$P$_{3/2}$ &
\dec 48398.941\tablenotemark[1] & -1267 & -9~~ \\
     &       &        &                & $3d^44p$ & $ ^6$P$_{5/2}$ &
\dec 48491.105\tablenotemark[1] & -1168 & -16~~ \\
     &       &        &                & $3d^44p$ & $ ^6$P$_{7/2}$ &
\dec 48632.125\tablenotemark[1] & -1030 & -13~~ \\

 ~25 & Mn~II & $3d^54s $ & $ ^7$S$_3$ & $3d^54p $ & $ ^7$P$_2$ & 
\dec 38366.184\tablenotemark[1] & 918 & 34~~ \\
    & & &            & $3d^54p $ & $ ^7$P$_3$ & 
\dec 38543.086\tablenotemark[1] & 1110 & 19~~ \\
    & & &            & $3d^54p $ & $ ^7$P$_4$ & 
\dec 38806.664\tablenotemark[1] & 1366 & 27~~ \\

 ~26 & Fe~II & $3d^64s $ & $ ^6$D$_{9/2}$ & $3d^64p $ & $ ^6$D$^o_{9/2}$ & 
\dec 38458.9871\tablenotemark[5] & 1449 & 2~~ \\
 & & &       & $3d^64p $ & $ ^6$D$^o_{7/2}$ & 
\dec 38660.0494\tablenotemark[5] & 1687 & -36~~ \\
 & & &       & $3d^64p $ & $ ^6$F$ _{11/2}$ & 
\dec 41968.0642\tablenotemark[5] & 1580 & 29~~ \\
 & & &       & $3d^64p $ & $ ^6$F$  _{9/2}$ & 
\dec 42114.8329\tablenotemark[5] & 1730 & 26~~ \\
 & & &       & $3d^64p $ & $ ^6$F$  _{7/2}$ & 
\dec 42237.0500\tablenotemark[5] & 1852 & 26~~ \\
 & & &       & $3d^64p $ & $ ^6$P$  _{7/2}$ & 
\dec 42658.2404\tablenotemark[5] & 1325 & 47~~ \\
 ~30 & Zn~II & $3d^{10}4s $ & $ ^2$S$_{1/2}$ & $3d^{10}4p $ & $ ^2$P$_{1/2}$ & 
\dec 48480.992\tablenotemark[1] & 1445 & 66~~ \\
 & & &         & $3d^{10}4p $ & $ ^2$P$_{3/2}$ & 
\dec 49355.027\tablenotemark[1] & 2291 & 94~~ \\
 ~32 & Ge~II & $4s^24p $ & $ ^2$P$_{1/2}$ & $4s^25s $ & $ ^2$S$_{1/2}$ & 
\dec 62403.027\tablenotemark[1] & -575 & -16~~ \\
\end{tabular}
\tablenotetext[1]{Morton, Ref. \cite{Morton}.}
\tablenotetext[2]{Moore, Ref. \cite{Moore}.}
\tablenotetext[3]{Pickering, Thorne and Webb, Ref. \cite{MgII}.}
\tablenotetext[4]{Sugar and Coliss, Ref. \cite{Sugar}.}
\tablenotetext[5]{Nave {\it et al}, Ref. \cite{Fe}.}
\end{table}

\begin{table}
\caption{Relativistic shift of energies of some methastable states of atoms
which used as optical frequency standards (cm$^{-1}$).}
\label{tabFS}
\begin{tabular}{cllllllrr}
 ~Z & Atom/Ion & \multicolumn{2}{c}{Ground state} & 
\multicolumn{2}{c}{Upper states} & $\omega_0$ & $q_1$ & $q_2$~~ \\
\hline
 ~20 & Ca~I & $4s^2 $ & $ ^1$S$_0$ & $4s4p $ & $ ^3$P$_1$ & 
\dec 15210. & 230 & 0~~ \\

 ~38 & Sr~II & $5s$   & $ ^2$S$_{1/2}$ & $4d$ & $ ^2$D$_{3/2}$ &
\dec 14555.90 & 2636 & 96~~ \\
     &       &        &                & $4d$ & $ ^2$D$_{5/2}$ &
\dec 14836.24 & 2852 & 160~~ \\

 ~70 & Yb~II & $6s$   & $ ^2$S$_{1/2}$ & $5d$ & $ ^2$D$_{3/2}$ &
\dec 24332.69 & 9898 & 1342~~ \\
     &       &        &                & $5d$ & $ ^2$D$_{5/2}$ &
\dec 22960.80 & 8298 & 1570~~ \\

 ~80 & Hg~II & $5d^{10}6s $ & $ ^2$S$_{1/2}$ & $5d^96s^2 $ & $ ^2$D$_{5/2}$ & 
\dec 35514.0 & -36785 & -9943~~ \\
    &      &        &                       & $5d^96s^2 $ & $ ^2$D$_{3/2}$ & 
\dec 50552.0 & -19377 & -12313~~ \\
\end{tabular}
\end{table}

\end{document}